# Detection of Structural Change in Geographic Regions of Interest by Self Organized Mapping: Las Vegas City and Lake Mead across the Years

**John M. Wandeto** [1, 2], **Henry O. Nyongesa** [2], **Birgitta Dresp-Langley** [1,*]

[1] ICube UMR 7357 CNRS - University of Strasbourg, France; birgitta.dresp@unistra.fr
[2] Dedan Kimathi University of Technology, Nyeri, Kenya; john.wandeto@etu.unistra.fr

[*] Correspondence: birgitta.dresp@icube.unistra.fr ; Tel.: +00-338-811-9117

**Abstract:** Time-series of satellite images may reveal important data about changes in environmental conditions and natural or urban landscape structures that are of potential interest to citizens, historians, or policymakers. We applied a fast method of image analysis using Self Organized Maps (SOM) and, more specifically, the quantization error (QE), for the visualization of critical changes in satellite images of Las Vegas, generated across the years 1984-2008, a period of major restructuration of the urban landscape. As shown in our previous work, the QE from the SOM output is a reliable measure of variability in local image contents. In the present work, we use statistical trend analysis to show how the QE from SOM run on specific geographic regions of interest extracted from minimally preprocessed satellite images can be exploited for detecting the magnitude and the direction of structural change across time. Significantly correlated demographic data for the same reference time period are highlighted. The approach is fast, reliable, and can be implemented for the rapid detection of potentially critical changes in time series of large bodies of image data.

## 1. Introduction

The analysis of time series of images by computational methods or algorithms represents a complex challenge in science and society. The detection and characterization of critical changes in public spaces of the natural or the built environment reflected by changes in image time series such as photographs or remotely sensed image data may be of considerable importance for risk mitigation policies and public awareness. This places a premium on fast automatic techniques for discriminating between changed and unchanged contents in large image time series, and computational methods of change detection in image data including remotely sensed data, exploiting different types of transforms and algorithms, have been developed to meet this challenge. Existing methods have been reviewed previously in excellent papers by [1] and [2]. Known computations include Otsu's algorithm [3], Kapur's algorithm [4], and various other procedures such as pixel-based change detection, image differencing, automated thresholding, image rationing, regression analysis on image data, the least-square method for change detection, change vector analysis, median filtering, background filtering, and fuzzy logic algorithms [5-11]. The scope of any of these methods is limited by the specific goal pursued; for detailed reviews, see [1, 2].

In general terms, change detection consists of identifying differences in the state of an object or phenomenon by observing it at different times and implies being able to quantify change(s) due to the effect of time on that given object or phenomenon. In image change detection this involves being able to reveal critical changes through analysis of discrete data sets drawn from image time series. One of the major applications of change detection concerns remotely sensed data obtained from Earth-orbiting satellites. These provide image time series through repetitive coverage at short intervals with consistent image quality, as shown previously in [1]. In this study here, we use a fast, unsupervised change detection technique based on the functional principles of self-organizing maps (SOM), first introduced by Kohonen [12]. Extracts from satellite images representing specific geographic regions of interest of Las Vegas County were used as input to SOM. After preprocessing to ensure equivalence in scale and contrast intensity of the extracted images within a time series, the image input is exploited directly without additional or intermediate procedures of analysis, bridging a gap between classic machine learning and traditional methods of geographic image analysis [13, 14]. Our previous work [15] had shown that a specific output variable of the SOM, the quantization error (*QE*), can be exploited as a diagnostic indicator for the presence of potentially critical local changes in medical image contents. In the present work, we provide proof-of concept simulations showing that the QE in the SOM output is significantly sensitive to spatial extent and intensity of local contrast differences in images. To control for differences in intensity across images of a given time series, a transform is applied before running SOM on each image of the time series. Thereafter, the QE from the SOM output provides a statistically reliable indicator of changes in the spatial extent of contrast regions across image contents, as will be shown. We then use the *QE* output from SOM on adequately preprocessed extracts from satellite images of Las Vegas County generated across the years 1984-2008. The image extracts correspond to two distinct geographic regions of interest (ROI) here: Las Vegas City Center and Lake Mead and its close surroundings in the Nevada Desert. The reference time period chosen for this study here (1984-2008) is of particular interest because of 1) major structural changes in the urban landscape of Las Vegas City during that period, and 2) the gradual dwindling of Lake Mead's water levels due to the effects of global climate change. We use statistical trend analysis to prove that the QE from the SOM on the different image ROI reliably reflects these critical changes across the years. Using Pearson's correlation analysis, we show that the QE output is significantly correlated with the most relevant demographic data for the same reference time period.

## 2. Results

This section is divided into three parts. In the first, results from a set of proof of concept studies are shown, with SOM run on monochromatic images with increasing spatial extent of contrast at constant intensity (first series of six images), and on images with increasing contrast intensity at constant spatial extent of contrast (second series of six images). This part is to show the statistically significant sensitivity of the QE to the spatial extent and the intensity of contrast in image series. In the first and the third parts, QE results from SOM run on the time series of images for the two geographic ROI are shown and discussed in the light of their statistical significance from linear trend analysis. Pearson's correlation method is used to link the QE to a set of highly relevant demographic data from the same reference time period. Image extraction and preprocessing applied before SOM analysis, and the principles of SOM run on the different image series are explained, with illustrations, in the Materials and Methods section. The Python code for the SOM is provided in the Supplementary Materials section [S1]. The 25 preprocessed images of the time series for each of the two ROI, Las Vegas City Center and Lake Mead, on which SOM analyses were run, are also provided [S2, S3].

*2.1. Sensitivity of the QE to spatial extent and intensity of contrast: proof of concept*

Results from four-by-four SOM run on each of the 12 test images for the proof of concept (see Materials and Methods) produced quantization errors (QE) that reveal a statistically signficant sensitivity to the spatial extent (Figure 1) and the intensity (Figure 2) of contrast in the sample images used for the proof of concept. Copies of these images are shown for illustration in the Figures 12 and 13 of the Materials and Methods section.

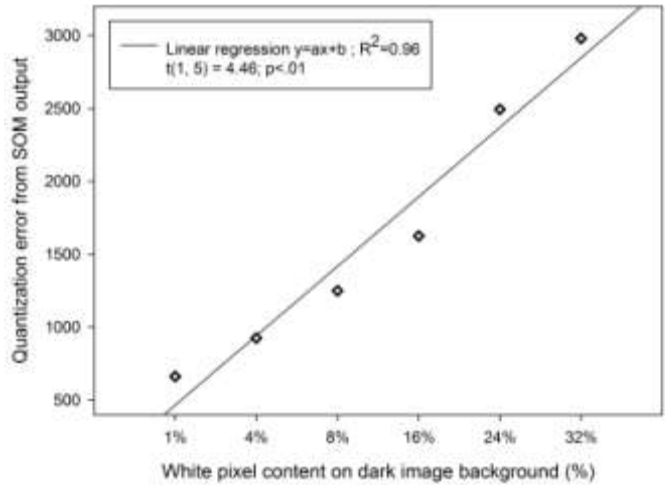

**Figure 1**: **Variations in QE output plotted as a function of spatial extent of contrast.** QE values are taken from the SOM on each of the six images of Figure 12 described in Materials and methods. When the intensity of contrast is, as here in the six test images, constant across images, the QE is shown to increase linearly with the increase in spatial extent of contrast, expressed here in % of the total image area, across the six images.

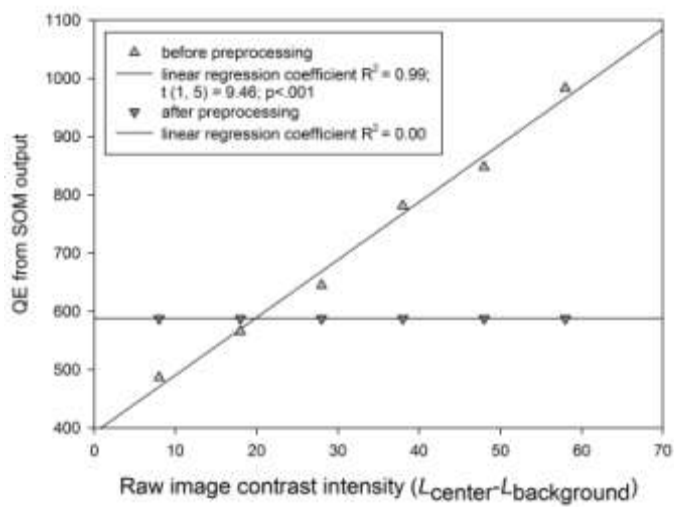

**Figure 2**: **Variations in QE output plotted as a function of contrast intensity.** QE values are taken from the SOM on each of the six images of Figure 13 described in Materials and Methods. When the spatial extent of contrast is constant, as in the six test images here, across images, the QE is shown to increase linearly with the intensity of contrast, expressed here in terms of the difference in luminance between the central image surface and its background.

Trend analysis (linear regression analysis) on the data in Figure 1 reveals a significant linear relationship between the magnitude of the QE and the spatial extent of contrast, with a linear correlation coefficient $R^2$ = .965 and a $t$ (1, 5) = 6.86, significant at $p<.01$. The correlation coefficient $R^2$ results from the least-squares fitting process of the variance of the residuals to the variance of the dependent variable and represents 1 *minus* the *ratio* of that variance. $R^2$ gives an estimate of the fraction of variance in the data that is explained by the fitted trend line. The statistical significance of the trend is determined by the statistical probability that the linear trend is significantly different from zero using the Student distribution (*t*) with *n*-1 degrees of freedom (DF). Linear regression analysis on the data in Figure 2 reveals also a close-to-perfect linear relationship between the magnitude of the QE and the intensity of contrast, with a linear correlation coefficient ($R^2$) of .99 and a *t* (1, 5) of 9.46, significant at p<.001. The dependency of the QE on contrast intensity is shown to disappear (as indicated by the straight line in the graph shown in Figure 2 here) after preprocessing of the images using a contrast normalization transform, applied here to all images before running SOM, as explained in the Materials and Methods section.

*2.2. Region of interest: Las Vegas City*

This section deals with the results from SOM run after pre-processing (see Materials and Methods) on satellite image extracts of the geographic ROI Las Vegas City. The corresponding image ROI covers an area of 72 by 98 pixels. The 25 extracted and pre-processed images are made available in the Supplementary Materials section [S2]. During the reference time period of this study, significant changes in building density across the whole of Las Vegas City has marked these years of major restructuration (1984-2009). Photographic snapshots taken of parts of the "Strip", the central artery of Las Vegas, where most attractions are located, illustrate some of these changes (Fig 3).

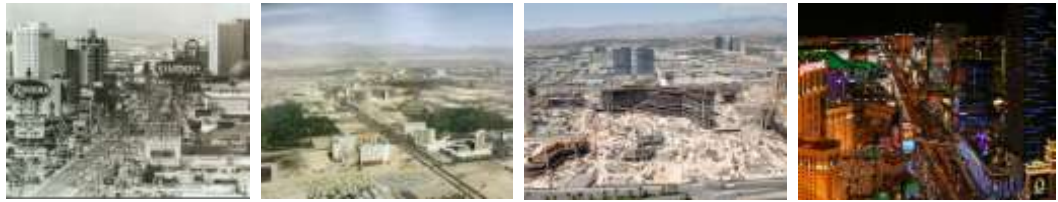

**Figure 3: Photographic snapshots of parts of "The Strip" across the years 1982-2010.** These photos, taken in 1984, 1995, 2005 and 2008 respectively, give some idea of the structural changes that took place in Las Vegas City during the reference time period of this study here, for which the satellite images generated by NASA were retrieved (see Materials and Methods).

Results in terms of QE output of the SOMs for the 25 image extracts corresponding to the ROI Las Vegas City are shown here below in Figure 4 as a function of the year in which the image was taken. The variations in the QE shown reflect varying spatial contrast distribution in the images across time. These are indicative of the major structural changes in the urban landscape during the reference time period. Trend analysis (linear regression analysis) on the data in Figure 4 reveals a trend towards increase in QE as a function of time expressed in terms of the progression in years of the reference period 1984-2008, with a linear regression coefficient $R^2$ = .48, and a *t* (1, 24) = 2.13, p<.05 indicating that the trend towards increase in QE with time is statistically significant. Contrast intensity across images of a time series for a given ROI being controlled for by preprocessing (see Materials and Methods), the significant increase in QE from SOM output reliably signals a significant increase in spatial extent of contrast regions in the images with time. Trend analysis on related demographic data for Las Vegas City show that both the number of visitors in millions and the estimated total population in thousands have increased significantly over the reference time period. These data are shown in the Figures 5 and 6 here below.

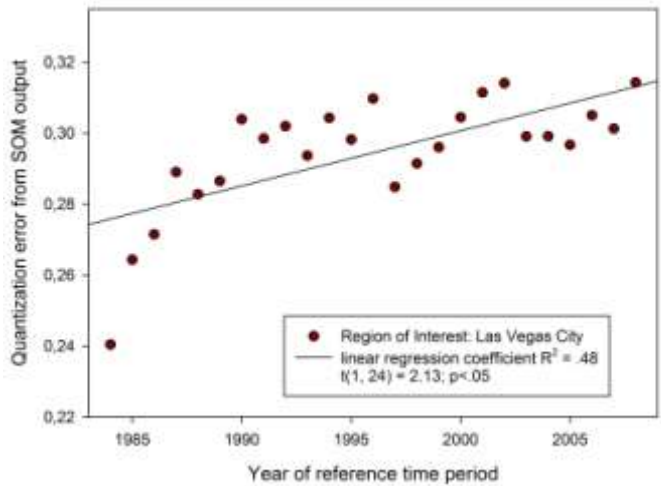

**Figure 4**: **Variations in QE output plotted as a function of time.** QE values are taken from the SOM on each of the 24 images corresponding to the geographic ROI 'Las Vegas City' (see Materials and Methods).

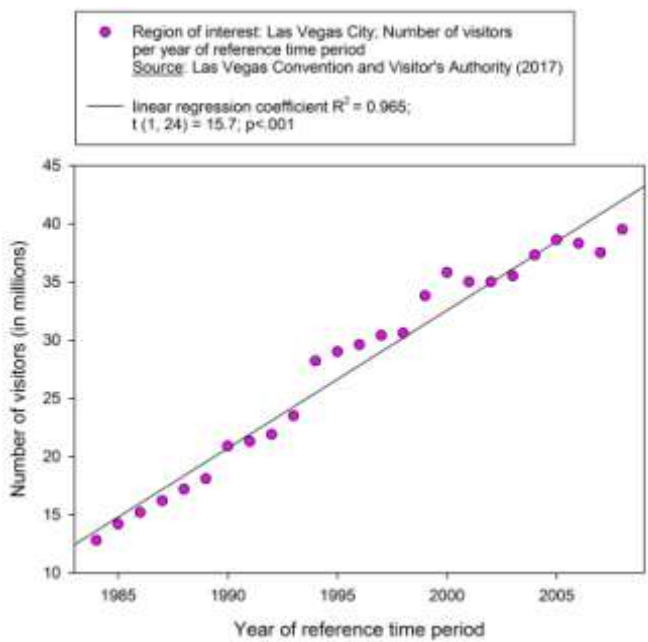

**Figure 5**: **Number of visitors of Las Vegas City as a function of time.** The linear trend yields a regression coefficient $R^2$ = .965 and is statistically significant with $t$ (1, 24) = 15.7, p<.001.

To assess the statistical correlation between the QE and other relevant demographic variables from the reference time period here, we computed Pearson's correlation coefficient on the distributions for *QE vs number of visitors* and *QE vs estimated population total*. The results of these analyses are shown in figures 7 and 8 here below. Pearson's correlation coefficient *R* gives an estimate of the statistical relationship, or association, between two independent continuous phenomena, or variables based on the mathematical concept of covariance. *R* is associated with a probability *p* and carries information about the magnitude of the association, or correlation, as well as the direction of the relationship. Pearson's correlation statistic computed on the paired distributions signals statistically significant correlations between *QE* and *number of visitors* (see Fig 6) and between *QE* and *population totals* (see Fig 7) for the time period 1984-2008 on the ROI Las Vegas City.

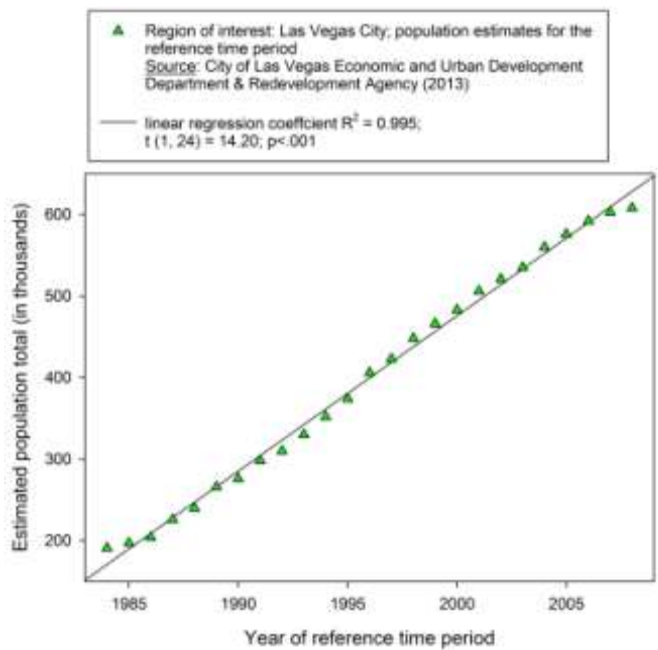

**Figure 6**: Estimated population total **for Las Vegas as a function of time.** The linear trend yields a regression coefficient $R^2$ = .995 and is statistically significant with $t$ (1, 24) = 14.2, p<.001.

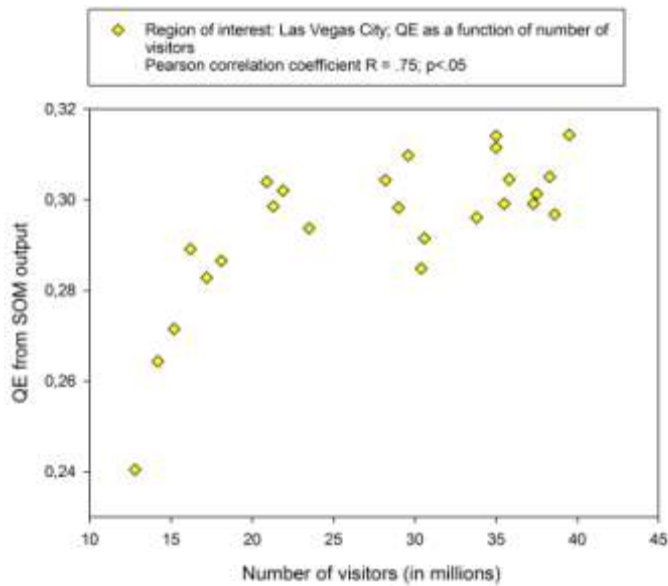

**Figure 7**: **Variations in QE output plotted as a function of the number of visitors of Las Vegas City during the reference time period.** Pearson's correlation statistic computed on paired distributions, plotted here in ascending order, gives a statistically significant correlation with $R$ = .75, $p$<.05.

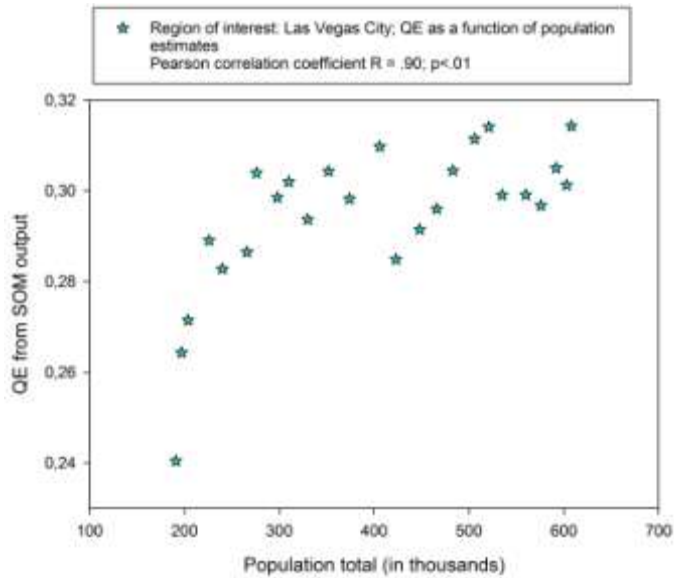

**Figure 8**: **Variations in QE output plotted as a function of the number of estimated population totals for Las Vegas City during the reference time period.** Pearson's correlation statistic computed on the paired distributions, plotted here in ascending order, gives a statistically significant correlation with *R* = .90, *p*<.01.

*2.3. Region of interest: Lake Mead Reservoir*

This section deals with the results from SOM run after pre-processing (see Materials and Methods) on satellite image extracts of the geographic ROI Lake Mead. The corresponding image ROI covers an area of 430 by 366 pixels. The 25 extracted and pre-processed images are made available in the Supplementary Materials section [S3]. Lake Mead is an artificial lake in the Nevada Desert collecting water from the Colorado River. Enclosed by Hoover Dam, the lake constitutes a reservoir serving water to the states of Arizona, California, and Nevada. Providing for nearly 20 million people and large areas of farmland, it is geographically situated 24 miles away from the Las Vegas Strip, southeast of Las Vegas City, Nevada, in the states of Nevada and Arizona. Lake Mead is the largest reservoir in the United States in terms of water capacity. During the reference time period of this study, Lake Mead water levels have progressively dwindled away as a consequence of global climate change. This phenomenon is captured by the QE from SOM run on the satellite image extracts, as shown here below in Figure 9.

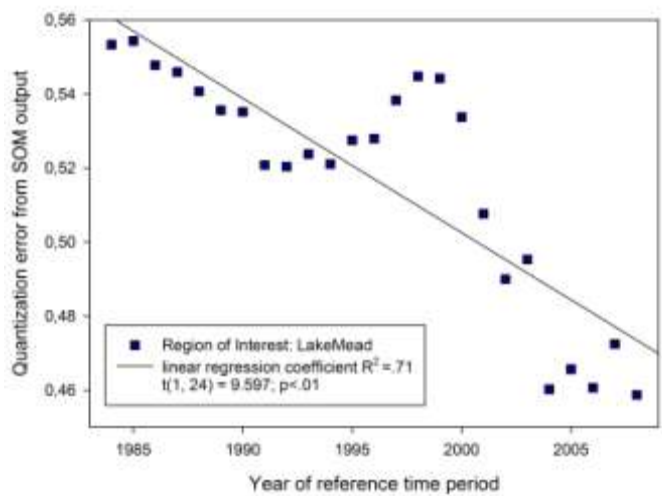

**Figure 9**: **Variations in QE output plotted as a function of time.** From SOM on image extracts for the ROI Lake Mead (see Materials and Methods).

Trend analysis (linear regression analysis) on the data in Figure 9 reveals a trend towards decrease in QE as a function of time expressed in terms of the progression in years of the reference period 1984-2008, with a linear regression coefficient $R^2 = .71$, and a $t(1, 24) = 9.597$, $p<.01$ indicating that the trend towards decrease in QE with time is statistically significant. Contrast intensity across images of a time series for a given ROI being controlled for by preprocessing (see Materials and Methods), the significant decrease in QE from SOM output reliably signals a significant decrease in spatial extent of contrast regions in the images with time. This image phenomenon is directly related to the shrinking away of Lake Mead, as can be seen in the image examples for 1984 and 2009 (see Materials and Methods). Water level statistics for the reference time period are shown here below in Figure 10.

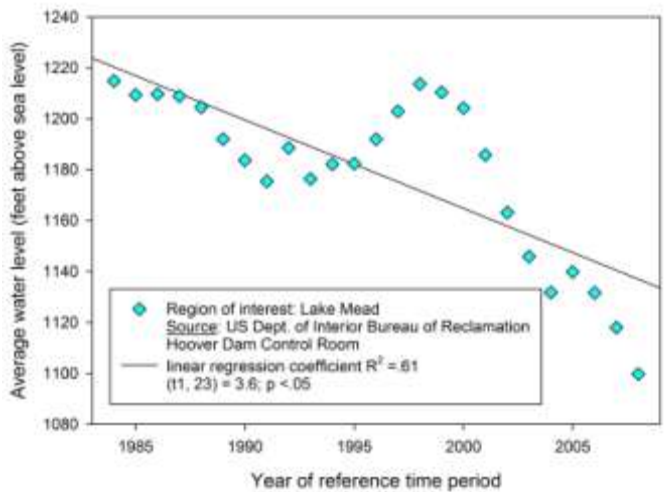

**Figure 10**: W**ater levels of Lake Mead as a function of time.** The linear trend towards decrease yields a regression coefficient $R^2 = .36$ and is statistically significant with $t(1, 24) = 15.7$, $p<.05$.

Pearson's correlation statistic computed on the paired distributions for *QE* and *water levels* signals a statistically significant correlation, shown here below in Figure 11, for the time period 1984-2008 on the ROI Lake Mead.

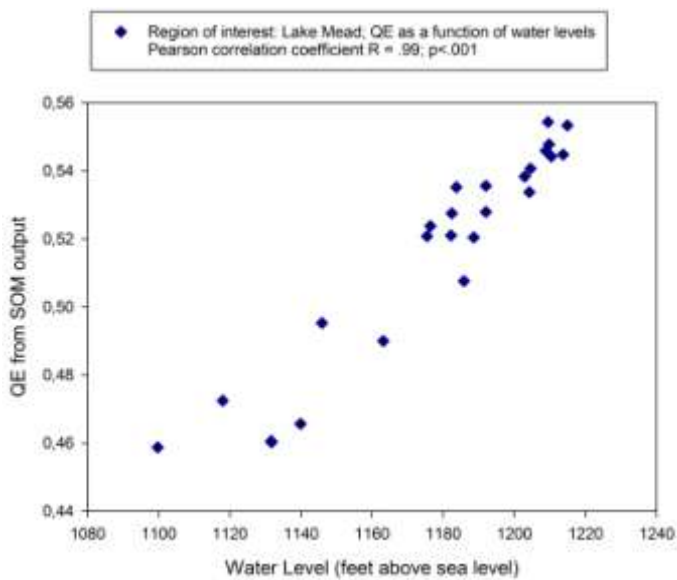

**Figure 11**: **Variations in QE output plotted as a function of the water levels of Lake Mead during the reference time period.** Pearson's correlation statistic computed on the paired distributions, plotted here in ascending order, gives a statistically significant correlation with $R = .99$, $p<.001$.

## 3. Discussion

In our previous work [15], we have shown that the quantization error ($QE$) of the output of image analysis by Self Organized Mapping (SOM), an unsupervised neural learning algorithm, is a reliable indicator of small local changes (reflecting, for example, tumor growth) in medical images that are not detectable by human vision. Here we provide further proof of concept by showing that the QE is sensitive to spatial extent of local image contrasts at constant intensity, and to intensity of contrasts of constant spatial extent. Applied here to adequately preprocessed extracts from satellite images, the QE output from SOM reliably reflects spatial variability and provides a statistically significant indicator of local changes in image contents in time reflecting the potentially critical evolution of man-induced and natural phenomena in geographic regions of interest. A major advantage of the method is the fast computation time. After some minimal preprocessing to control for equivalence in spatial scale and contrast intensity of images of a given time series, it takes less than a minute to run SOM on a time series of 25 images. SOM analysis on time series of satellite images is easy to implement [21, 22, 23], fast, and represents a promising technique for the automatic tracking and harvesting of landscape information from large bodies of image data, providing inexpensive, ready-to-use, and reliable simulations.

## 4. Materials and Methods

The images used for the proof-of-concept simulations showing the sensitivity of the QE from SOM output to spatial extent of contrast at constant intensity (Figure 12) and to contrast intensity at constant spatial extent of contrast (Figure 13) are shown here below.

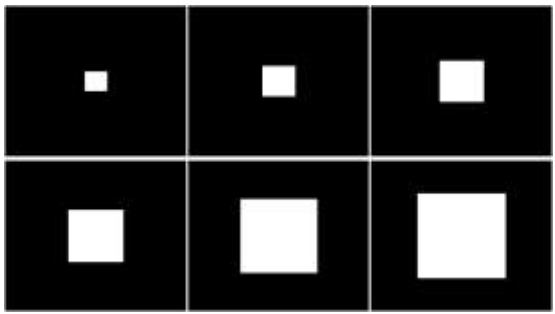

**Figure 12: Six test images for the proof-of-concept SOM testing for QE sensitivity to spatial extent of contrast in images.** Identically sized images with increasing spatial extent of contrast across images at a constant contrast intensity of 60 cd/m$^2$ on a background of 2 cd/m$^2$ were generated for these simulations.

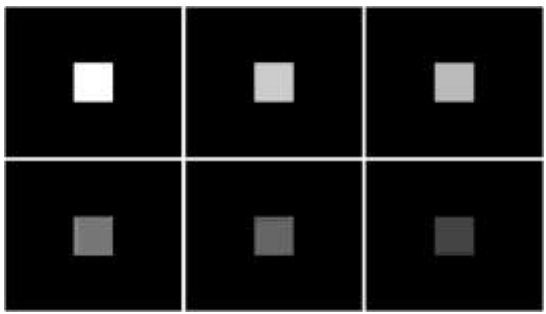

**Figure 13: Six test images for the proof-of-concept SOM testing for QE sensitivity to the intensity of contrast in images.** Identically sized images with gradually decreasing/increasing contrast intensity of constant spatial extent across images were generated for these simulations.

The last image of each series was used to train a 4 by 4 SOM with a neighborhood distance of 1.2 and a learning rate of 0.2 for 10,000 iterations. The QE from the SOM analyses run on each of the different images was fed into linear regression analysis (see Results).

The images used for analysis of geographic regions of interest by SOM were extracted from time-lapse animations of Las Vegas City and Lake Mead, Nevada, from 1984-2008, as captured by NASA Landsat sensors [16]. VLC, an open source media player, was used to extract the images from the time-lapse animations provided. The images are false-color, showing healthy vegetation in red while water is in black. Samples of the 25 images extracted for each of the two geographic ROI, Las Vegas City and Lake Mead, are shown in Figure 14 here below.

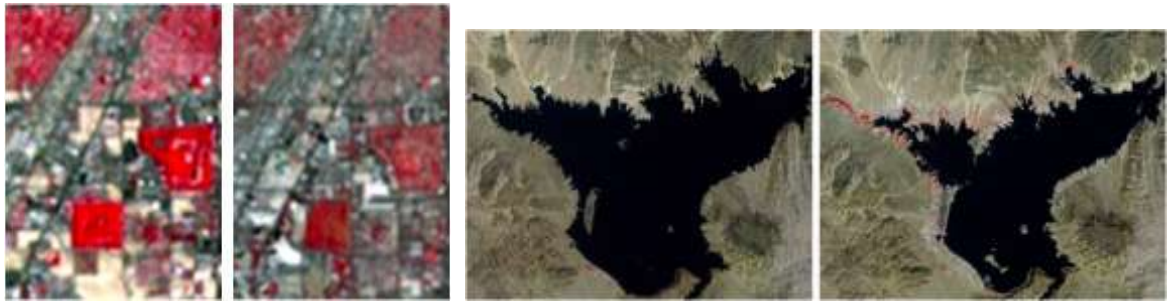

**Figure 14**: **Examples of images extracted from the NASA Landsat database showing two geographic regions of interest.** The pre-processed images of Las Vegas City from the years 1984 and 2008 are shown on the left, pre-processed images of the region around Lake Mead from years 1984 and 2008 are shown on the right.

Before running SOM on the time series for each geographic ROI, the images were pre-processed to ensure that images from a same subset are aligned. This was achieved by applying the method of co-registration [17] using *StackReg* [17], which is a plug-in for *ImageJ*, an open source image processing program designed for scientific multidimensional images. From each image series for the two geographic ROI, the last image was used to anchor the registration. Control for variations in contrast intensity between images of a series was performed after registration. This was achieved by increasing the image contrast and by removing any local variations at different times of image acquisition [18]. For each extracted image, contrast normalization was obtained using

$$I_{final} = (I - I_{min}/I_{max} - I_{min}) \times 255$$

The registered and normalized image taken in 2011 from each ROI was used to train a 4 by 4 SOM with a neighborhood distance of 1.2 and a learning rate of 0.2 for 10,000 iterations [20]. Since the original images used color to emphasize different areas on the maps, pixel-based RGB values are used as input features to the SOM. This ensures a pixel-by-pixel capture of detail and avoids errors due to inaccurate feature calculation which often occur with complex images [21]. With the trained SOM, QE values were determined for each of the 25 images of each geographic region of interest from the years 1984-2008. The QE values were fed into linear regression analysis (see Results). Correlation analysis using Pearson's method were performed to establish the statistical significance of correlations between variations in QE and other demographic data for the same reference time period (see Results).

**Acknowledgments:** This research was supported by a study grant from the French Embassy in Nairobi to J. W.